\documentclass[
  journal=largetwo,
  manuscript=article-type,
  year=2025,
  volume=37
]{cup-journal}

\usepackage{amsmath,amssymb,amsfonts}
\usepackage[nopatch]{microtype}
\usepackage{booktabs}
\usepackage{stackengine}
\usepackage{graphicx}
\usepackage{textcomp}
\usepackage{xcolor}
\usepackage{ulem} 
\usepackage{aastex_macros} 
\PassOptionsToPackage{hyphens}{url}\usepackage{hyperref} 

\title{Optimising the Processing and Storage of Visibilities using lossy compression}

\author{R. Dodson} 
\affiliation{{Int. Centre for Radio Astronomy Research},  {University of Western Australia},
Perth, Australia}
\email[R. Dodson]{richard.dodson@icrar.org}

\author{A. Williamson} 
\affiliation{{Int. Centre for Radio Astronomy Research}, {University of Western Australia},
Perth, Australia}
\email[A. Williamson]{alex.williamson@icrar.org}

\author{Q. Gong} 
\affiliation{{Oak Ridge National Laboratory}, Oak Ridge, USA}
\email[Q. Gong]{gongq@ornl.gov}

\author{P. J. Elahi} 
\affiliation{{Pawsey Supercomputing Research Centre}, Perth, Australia}
\email[P. J. Elahi]{Pascal.Elahi@csiro.au}

\author{A. Wicenec} 
\affiliation{{Int. Centre for Radio Astronomy Research}, {University of Western Australia}, Perth, Australia}
\email[A. Wicenec]{andreas.wicenec@icrar.org}

\author{M. J. Rioja} 
\affiliation{{Int. Centre for Radio Astronomy Research}, {University of Western Australia}, Perth, Australia}
\alsoaffiliation{Observatorio Astron\'omico Nacional (IGN), Alfonso XII, 3 y 5, 28014 Madrid, Spain}
\email[M. J. Rioja]{maria.rioja@icrar.org}

\author{J. Chen} 
\affiliation{{University of Alabama at Birmingham}, Birmingham, USA}
\email{jchen3@uab.edu}

\author{N. Podhorszki} 
\affiliation{{Oak Ridge National Laboratory}, Oak Ridge, USA}
\email{pnorbert@ornl.gov}

\author{S. Klasky} 
\affiliation{{Oak Ridge National Laboratory}, Oak Ridge, USA}
\email{klasky@ornl.gov}

\author{{M. Meyer}} 
\affiliation{{Int. Centre for Radio Astronomy Research}, {University of Western Australia}, Perth, Australia}
\email{martin.meyer@uwa.edu.au}

\email[R. Dodson]{richard.dodson@icrar.org}


\newcommand\arcdeg{\mbox{$^\circ$}}

\newcommand\farcs{\mbox{$.\!\!^{\prime\prime}$}}

\begin{document}

\begin{abstract}
{
The next-generation radio astronomy instruments are providing a massive increase in sensitivity and coverage, largely through increasing the number of stations in the array and the frequency span sampled. 
The two primary problems encountered when processing the resultant avalanche of data are the need for abundant storage and the constraints imposed by I/O, as I/O bandwidths drop significantly on cold storage such as tapes. 
An example of this is the data deluge expected from the SKA Telescopes of more than {60}\,PB per day, all to be stored on the buffer filesystem.
%
While compressing the data is an obvious solution, the impacts on the final data products are hard to predict. 

In this paper, we chose an error-controlled compressor -- MGARD -- and applied it to simulated SKA-Mid and real pathfinder visibility data, in noise-free and noise-dominated regimes. 
As the data has an implicit error level in the system temperature, using an error bound in compression provides a natural metric for compression. 
MGARD ensures the compression incurred errors adhere to the user-prescribed tolerance. To measure the degradation of images reconstructed using the lossy compressed data, we proposed a list of diagnostic measures, exploring
the trade-off between these error bounds and the corresponding compression ratios, as well as the impact on science quality derived from the lossy compressed data products through a series of experiments.

We studied the global and local impacts on the output images for continuum and spectral line examples. 
We found relative error bounds of as much as $10\%$, which provide compression ratios of about 20, have a limited impact on the continuum imaging as the increased noise is less than the image RMS,
whereas a $1\%$ error bound (compression ratio of 8) introduces an increase in noise of about an order of magnitude less than the image RMS. 
For extremely sensitive observations and for very precious data, we would recommend a $0.1\%$ error bound with compression ratios of about 4. These have noise impacts two orders of magnitude less than the image RMS levels. At these levels, the limits are due to instabilities in the deconvolution methods.
{We compared the results to the alternative compression tool DYSCO, in both the impacts on the images and in the relative flexibility. MGARD provides better compression for similar error bounds, and has a host of potentially powerful additional features.}
%
}
\end{abstract}
\section{Introduction}
\subsection{The Square Kilometre Array}
Radio Astronomy is currently undergoing a paradigm shift,
with the planning for many next-generation radio instruments, such as the Square Kilometre Array (SKA), the next-generation Very Large Array (ngVLA) and the next-generation Event Horizon Telescope (ngEHT).
All of these provide at least an order of magnitude increase in bandwidth and a few orders of magnitude in collecting area (and sensitivity) over current radio telescopes.
This enhancement will provide us with the opportunities to survey the radio sky in exquisite detail, such as detecting the signal from the epoch of reionization from when the first stars were born \citep{ska-aas-eor} and measuring the spectral signal from millions of galaxies \citep{ska-aas-bg}.

To achieve this significant advance in our understanding, the radio astronomy community must manage and process unprecedented volumes of data \citep{ska-quinn}.
In this paper we focus on the SKA, as Australia is a founding member of the intergovernmental organisation, but our results are applicable to all the coming infrastructure. 
The SKA will be constructed in Australia for frequencies spanning 50 to 350MHz (SKA-Low) and in South Africa for frequencies from 350MHz to 15GHz (SKA-Mid) \citep{ska-overview}.
Phase 1 of SKA-Low will have 512 stations with a diameter of 38m each. Phase 1 of SKA-Mid will consist of 197 parabolic dishes with 15m diameter. 
The final goal is to have a full square kilometre of collecting area for both arrays.
Construction has commenced and science verification will begin in 2027\footnote{https://www.skao.int/en/science-users/118/ska-telescope-specifications}.

Following this, data rates from each of the correlators will become $\sim$6\,TB/s, making storage one of the largest cost drivers for the SKA project. Due to limited storage, raw data will be temporally captured into a local buffer and must be processed within a specific period -- ranging from days to weeks -- before being permanently erased. 
Since observatory data are unreproducible and further future analysis may be necessary,  savings in storage will not only directly impact the project's operational budget but also allow more data to be stored long term.

\subsection{High-performance I/O and Data Compression}

Nearly all modern radio astronomy analysis is performed via {the Common Astronomy Software Applications (CASA) software package} \citep{casa},
which provides an ipython environment and a set of core data processing tasks and utilities.
The CASACore Table Data System storage manager can use the Adaptable Input Output System version 2 (ADIOS2) \citep{WANG:2016, Godoy:2020} as the input/output (I/O) and storage backend. ADIOS is a software framework with a simple I/O abstraction and a self-describing data model centred around distributed data arrays, allowing multiple applications to publish and subscribe data at large levels of concurrency. 
It is primarily focused on high-performance, parallel I/O, with its parallel storage performance,
the file format, the memory management, and data aggregation algorithms being designed together to be highly scalable in every axis (many processes, many variables, large amounts of data, many output steps). 
{For the ADIOS/MGARD compression of the {Deep Investigation of Neutral Gas Origins} (DINGO) uv-gridded data \citep{williamson-24} we found a seven-fold reduction in the storage footprint and a seven-fold improvement on the processing speed.}
ADIOS2 additionally provides users the access to state-of-the-art lossless and lossy compressors (i.e. data is fully recovered after decompression, or the data is only approximately recovered) through its operator. By attaching the operator to a variable, ADIOS seamlessly implements the compression and I/O as a combined operation. 
One example of this is the use of the MGARD library of functions, which we are currently testing as an extension of the software for the {Australian SKA Pathfinder} (ASKAP), ASKAPSoft \citep{askapsoft}.

MGARD \citep{Gong:2023} is a software that offers error-controlled lossy compression rooted in multi-grid theories. It transforms floating-point scientific data into a set of multilevel coefficients through multi-linear interpolation and $L^2$ projection, followed by linear quantisation and lossless encoding processes. The magnitude of the transformed coefficients is close to zero, making them more amenable to compression than the original data.
{MGARD has been designed to provide error-controlled rather than fixed-bit compression. It guarantees the compression incurred errors to stay below user-prescribed error bounds, such that the lossy compressed data can be trusted for scientific usage. The resulted compression ratios are data dependent with smoother data being more compressible than noisy data.} 
One of MGARD's notable features is its array of error control options, including the various euclidean norms of $L^\infty$, $L^2$, point-wise relative $L^\infty$, and the ability to vary error bounds across regions or different frequency components. This flexibility is valuable for preserving Region-of-Interest (RoI) and/or Quantities-of-Interest (QoI) \citep{Gong:2022} derived from the reconstructed data. Though we do not explore the use of RoI and QoI in this paper, compression of astronomy data could be significantly improved in size and quality using these features; this is left for future work.  
For example, RoIs could be used to implement an optimal Baseline Dependent Averaging (BDA) approach on the visibilities.

MGARD has been optimised with highly-tuned CPU and GPU kernels and efficient memory and device management mechanisms, ensuring rapid operations and device portability. 
When integrated with ADIOS2, variables and the desired error bounds can be prescribed through ADIOS2's operator API, resulting in a self-describing compressed buffer containing all necessary parameters for decompression. 

The natural point of comparison for MGARD is the bit-reduction compression method {DYSCO (DYnamical Statistical COmpression)}, a lossy compressor specifically designed for radio astronomical data \citep{dysco}. 
DYSCO normalises the data across different antennas, polarisation, timesteps, and frequencies, ensuring a constant noise variance across the full dataset. 
It then performs non-linear quantisation followed by customised encoding. 
Unlike MGARD, DYSCO cannot directly prescribe error bounds; the compression-induced errors can only be confirmed post-factum, and its choice of quantisation bins is subjective to the type of normalisation. 
Previous literature indicates that the main benefit of using DYSCO is that its compression noise does not exhibit spatial structure. In this paper, we demonstrate that MGARD-compressed data exhibit the same properties as well as the previously mentioned advantages. 

Furthermore, although not investigated in this paper, MGARD is natively embedded in ADIOS, thus allowing fully parallel I/O just by linking to the relevant library, and MGARD is GPU-enabled, which improves the speed of the compression calculations. These advantages of MGARD are discussed in \citet{williamson-24}, \citet[in prep]{williamson-25} and future publications.

\subsection{Radio Astronomy Data}
\begin{figure}
    \centering
    \def\big{\includegraphics[height=7cm]{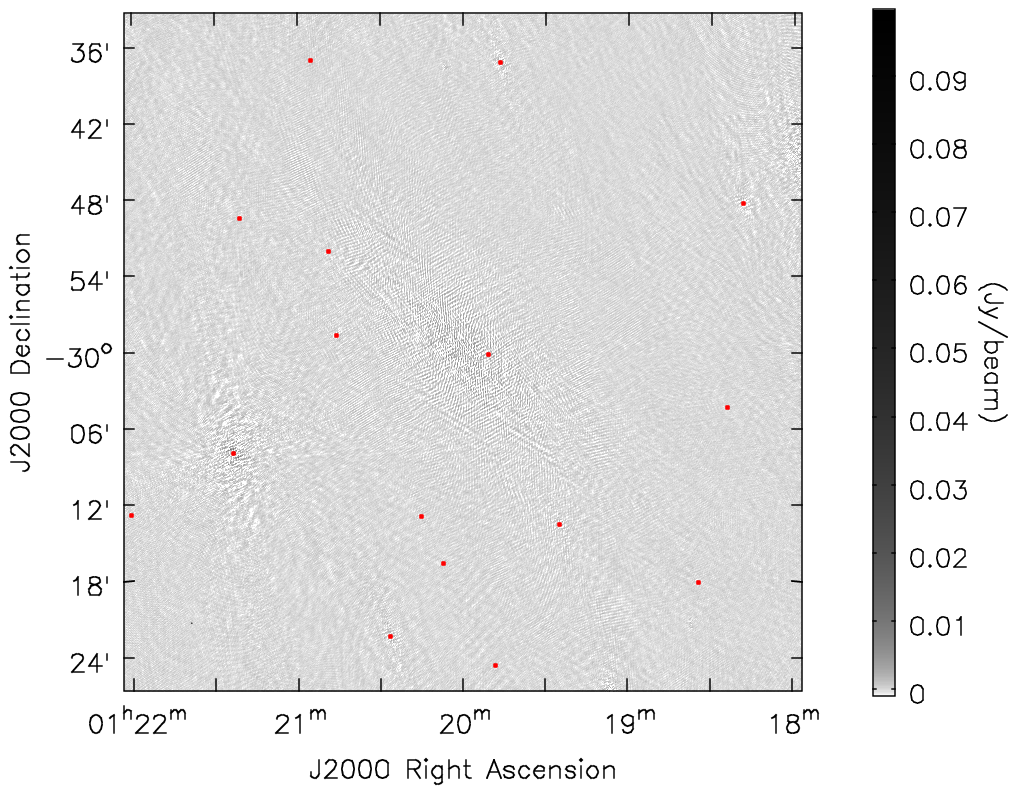}}
\def\little{\includegraphics[height=4cm]{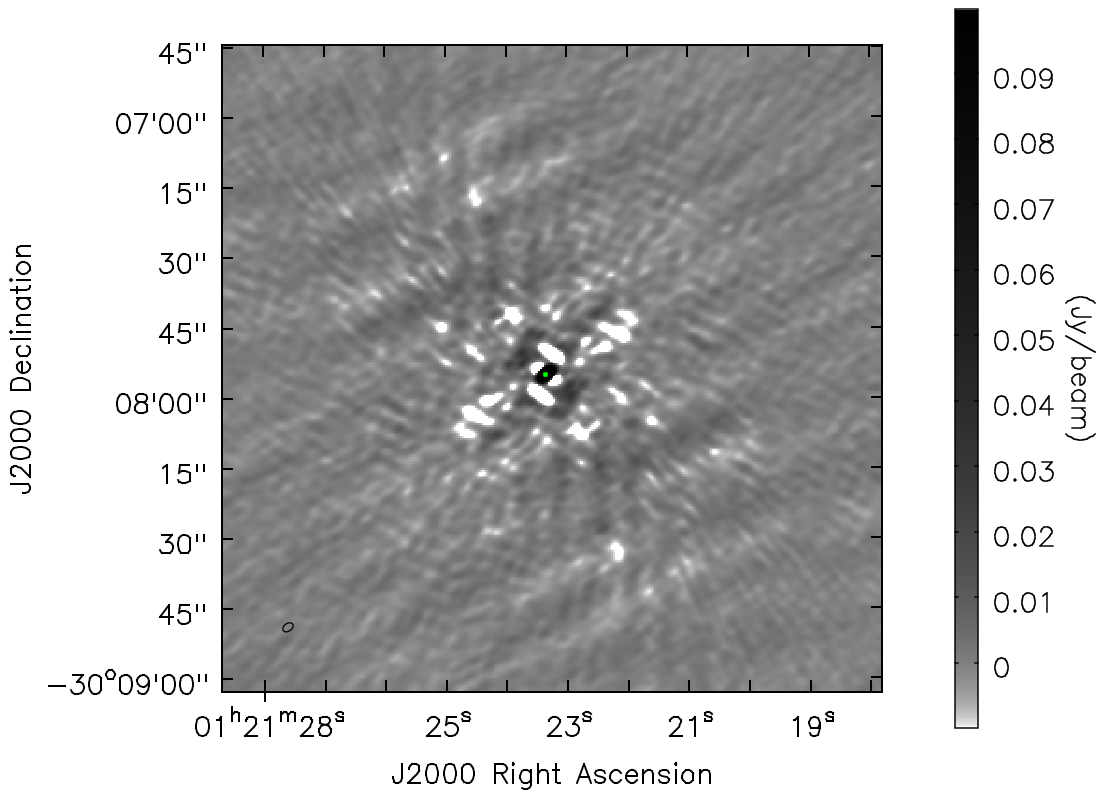}}
\stackinset{l}{-20pt}{t}{-20pt}{\little}{\big}

    \caption{Image of the data used in these investigations, based on the GLEAM catalogue to provide a realistic complex sky with added random thermal noise, and deconvolved with WSClean. The locations of in-field GLEAM-model components are marked with red squares and are widely dispersed over the image with only thermal noise between those regions of interest. 
    The insert shows the low flux residuals around the brightest source in the image of 1~Jy. This emphasises the interplay of the thermal noise and the imperfect reconstruction from the deconvolution, which sets a limit on the accuracy of reconstruction.
    }
    \label{fig:skyimage}
\end{figure}

Radio data presents a unique challenge.  Much of the data is originating from thermal noise as the portion of the sky containing emission is small, see for example Fig.~\ref{fig:skyimage} that shows the simulated sky used in these investigations, which is based on the real {GaLactic and Extragalactic All-sky MWA} (GLEAM) radio survey catalogue. 
In this figure, only a small fraction of pixels in the image contain emission from astronomical sources, which appear as spatially concentrated regions of high radio emission. Not all astronomical sources are spatially concentrated and with ever improving resolution, what was once a single source can be resolved into spatially extended, diffuse emission. Moreover, some signals, such as the sought-after signal of reionisation from the first stars, will be distributed across the entire image and is hidden in the noise \citep{Liu:2014,Nasirudin:2020}.

\begin{figure*}
    \centering
    \includegraphics[width=0.95\textwidth]{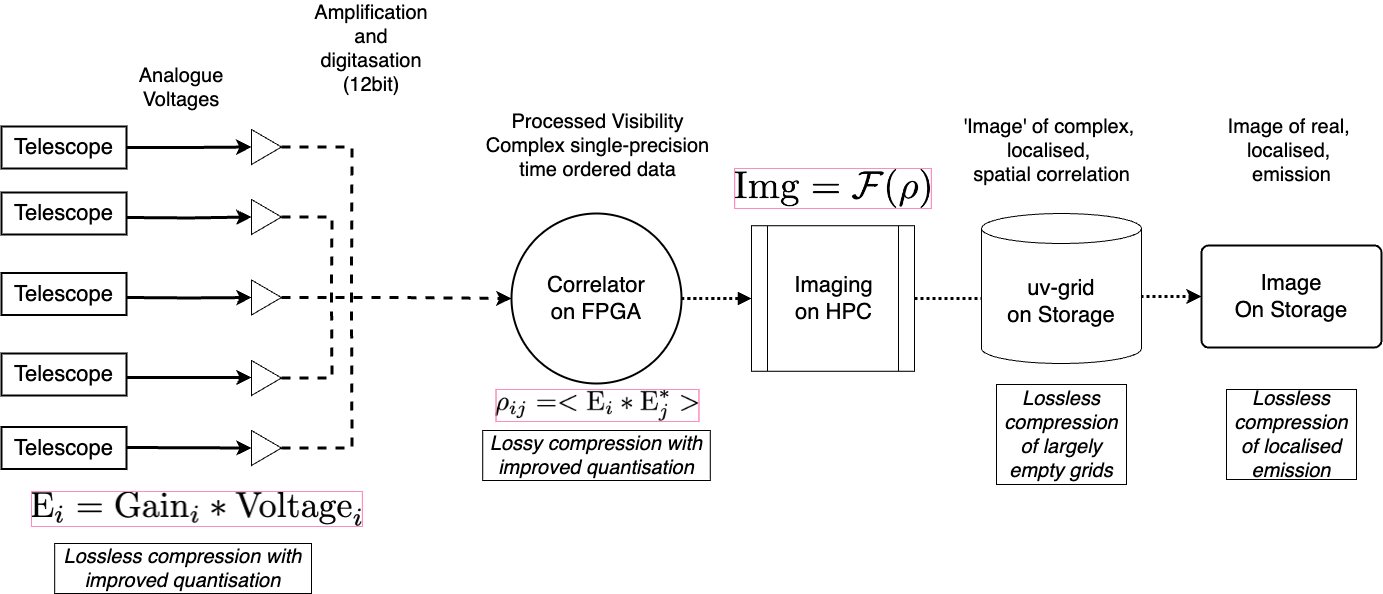}
    \caption{Data Flow in a typical radio interferometre from the individual telescope, stations through the correlator and imaging, to intermediate and final data products. Various datatypes and possible compression of these datatypes are noted along the path. 
    After the amplification and digitisation the data is limited to 12bits, but may not fully inhabit the data range. 
    After correlation the time-sampled data is integrated into a complex value, which is not compatible with lossless compression but can undergo lossy compression. 
    This is because the data should be thermally limited in accuracy (from the amplification) and this would normally be less than the nominal numerical precision. 
    For the intermediate data product of the temporal data resampled onto a regular spatial uv-grid, many of the cells will be empty and lossless compression will be effective. 
    In the final image the astronomical emission will be concentrated into limited regions of interest, allowing for some of the advanced features of MGARD to be applied.
    }
    \label{fig:dataflow}
\end{figure*}

Furthermore, the sky signals are collected in the Fourier domain, which is the reciprocal of the sky image domain. Thus the weak individual signals from the different sources are spread over the Fourier terms and the individual samples become completely dominated by the system noise. The latter should, in a good design, be limited by the thermal noise from the amplifier chain in the receivers. 
Figure \ref{fig:dataflow} itemises four of the places where data compression could be applied: on the input digital voltages, on the time-ordered visibilities, on the spatially gridded visibilities and on the final image. We are focusing only on the specifically Radio Astronomical domains of use. In this paper we investigate the compression of correlated outputs, in \citep[in prep]{williamson-25} we will present the results for spatially gridded uv-visibilities.

\begin{figure}
    \centering
    \includegraphics[width=0.99\textwidth, clip, trim={1.cm 0.cm 1.5cm 0.cm} ]{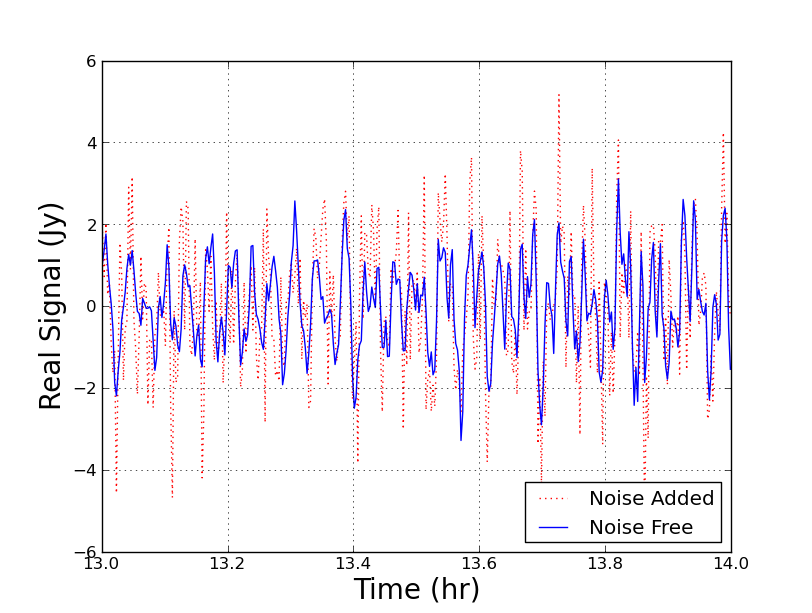}
    \caption{The real part of the signal on one baseline of length $\sim$16\,km, with and without noise, from the 100 channel complex sky model simulations where the noise and sky-signal are of comparable magnitude. 
    For the 1000 channel simulations the sky signal would be the same but the noise is $\sqrt{10}$ times greater,
    {thus this represents a low-noise, but not noise-free case. The majority of the analysis is on high-noise or noise-free cases.}
    }
    \label{fig:baseline}
\end{figure}

The raw samples from the correlator consist of a weak sky signal, which is precious and can not be distorted, and a strong random noise signal, which will be `averaged away' in the formation of the images of the sky. An example of the signals on a single baseline of about 16\,km is shown in Fig.~\ref{fig:baseline}, where the sky signal is from the GLEAM catalogue model, and an example of the noise levels from the system thermal contributions are included. 
{Thus, in most cases the noise dominates the signal, making the astronomical results equivalent to the noise-only results.}

This data analysis challenge is combined with a data volume challenge. Radio astronomy data volumes from current generation telescopes are of PB-scale. This data is often stored as a MeasurementSet (MS) \citep{kemball:2000}, a format in which visibility and single-dish data are stored to accommodate synthesis. Although this format has been historically very useful, it does not scale particularly well and often the science process requires non-optimal access, giving rise to additional I/O load. 
It is for these reasons that we are investigating the application of MGARD, implemented via the CASACore interface to ADIOS2 on the raw MS. In Radio Astronomy compression can be applied in multiple places along the data collection and processing chain: compression of images and multi-dimensional image cubes \citep{Kitaeff:2015}, particularly focusing on the sparseness of the radio sky to leverage the use of RoI \citep{Peters:2014}; compression of gridded data, where the sparsely filled sampling grid for the imaging can be compressed significantly in a lossless fashion \citep{williamson-24}; and relevant to this report, the lossy compression of the raw (correlator output) MS datasets. 

\section{Methods}
\subsection{Observational Data}

We used the SKA simulator for Radio Interferometry data, OSKAR \citep{oskar}, 
to simulate a clean 1 hour-long dataset based on the SKA-Mid AA2 (64 antenna) configuration\footnote{{https://gitlab.com/ska-telescope/sdp/ska-sdp-par-model/-/tree/add-AA-layouts/data/layouts}}(hereafter AA2-Mid) and a single polarisation, with the GLEAM catalogue \citep{gleam-cat} to provide 228 unpolarised sources within a $\sim$3 degree field of view. 
This represents a realistic complex sky, such as would be expected in real observations. This was done at 1.0GHz over a 300MHz bandwidth, with both 100 and 1000 channels. 
The GLEAM sky-model had a strongest source with a flux of just under 3\,Jy, and a standard deviation of about 1\,Jy. 
The baseline lengths with AA2-Mid range between 20\,m and 84\,km.
In addition we added a further column of data, consisting of the GLEAM sky-model plus pure normal-distributed Gaussian noise for each visibility.  
To represent the continuum case we added the expected per-baseline noise of 0.14\,Jy over the whole bandwidth, which was converted to the noise per visibility by scaling with the square root of the number of channels. Thus the thermal noise is several times greater than the sky signal. 
To investigate a simpler sky, where the sky signals are not varying so rapidly, we replaced the model with a single compact but slightly resolving source, using a few components at the centre of the field. This simple model had an integrated flux of 3\,Jy.
This was in order to achieve {very high dynamic ranges (greater than a thousand) with the limited uv-coverage of the 64 antenna AA2 configuration.}
To represent the spectral line case 
we added the simple model to a single channel of the GLEAM sky-model, scaled up by a factor of ten, so that it dominated the thermal noise.
This represented a strong, narrow, maser-like emission dataset. 
We imaged that channel and a few either side to test whether the compression introduces `bleed through' of apparent emission into other spectral channels. 

Our final test was to investigate compression on real data; we selected a single typical example of {an observation from the LOw-Frequency ARray (LOFAR)}. 
The data selected was observation ID L686982, heavily averaged (for data volume considerations), targeted on the European Large Area ISO Survey deep field N1 at 16:11:00 +54:57:00, averaged down to 230 channels of width 195kHz around 150MHz and 1\,minute integrations over the eight hour long observation.
The full scientific observations are published \citep{elais-n1} and although our heavily averaged version would not be suitable for a best-quality scientific image, it was suitable to test the behaviour of real data under compression (albeit with improved per visibility signal to noise).
With this real data we found that the data distribution was far from Gaussian, because of Radio Frequency Interference (RFI) and Not-a-Number (NaN) values. 
We truncated the data range to $\pm$100\,Jy with all other values flagged and set to zero. 
The standard deviation of the remaining data was 4\,Jy, albeit with a distribution that had an excess of values closer to zero and a long tail of large values. 
Nevertheless, we compressed it using MGARD in the same fashion as the simulated data, with our best estimate of suitable error bounds.


\subsection{MGARD Compression of complex visibility data}

The visibilities, being complex values, can be compressed separately in the real and imaginary form or in the amplitude and phase form; we trialled both presentations.
As compression algorithms work best on smoothly changing data, we also reorganised the native output array ordering from time-ordered (i.e. all cross-correlations on all baselines at every timestep) to baseline-ordered (i.e. each individual baseline in time, in sequence). This is a supported ordering in the MS format, but has the advantage of presenting any smoothly changing features, such as the signal amplitude, in a fashion most detectable for the compression algorithms. 

We have compressed the visibilities
using MGARD with relative error bounds (EB) between $6\times10^{-6}-5\times10^{-1}$, 
in steps of approximately a factor of 3.
Below this range the compression ratios approach unitary and above it one introduces appreciable distortions into the image.
The EB is defined as the global root-mean-square-error normalised by the range of data values. Where the range is positive to negative (as in the visibilities) an EB of 0.5 would represent the whole data span and the data could effectively be compressed to a single value for the whole dataset. 
{In practise, for EBs below 0.1, after MGARD compression the absolute error in the reconstructed data values forms a sharply truncated distribution between zero and a few times the error bound, with 99\% of the post-compression values below the requested EB.
Note that above 0.1 the simple relationship between the EB and the 99\% percentile starts to diverge.}
We tested the impact of data reordering -- in time-ordered and baseline-ordered changing fashion -- on compression ratios. 
{Finally, for the real data we used an absolute EB, i.e. bounding the compression incurred error by an absolute value rather than percentage relative to data value range.
}

\subsection{Comparison of MGARD Compression with DYSCO Compression}

The results were compared to similar analysis from data compressed with DYSCO, where the compression is limited to bit reduction. That is 32 bit float numbers were reduced to a representation with a lower quantisation; 2-, 3-, 4-, 6-, 8-, 10-, 12-, 16-bit are possible options. 
{After DYSCO compression the absolute error in the reconstructed data values forms a distribution between zero and a maximum value. 
The distribution does not truncate as sharply as for MGARD, but does allow us to measure an `error bound' in the cumulative probability function below which 99\% of the data is reconstructed sufficiently accurately.}
{DYSCO also uses a non-linear quantisation scheme, which means that the rarer numbers are less accurately stored. This is harder to compare directly, but we can test the impacts on the data quality estimates derived from the imaging.} 

{In practise, we found that DYSCO did not support the baseline-ordered data format, so for 
the relevant tests and comparisons all data was in the default time-ordered format and noise dominated.}
{We note that this places our comparison in the domain where DYSCO was reported to perform best; that is the low SNR domain.}

\subsection{Image Quality Analysis}
All the data was imaged with WSClean \citep{wsclean}, {with an image size of 8000$\times$8000 px$^{2}$}, a cell size of 0\farcs4, a taper of 2\farcs0 and 10,000 clean iterations. 
This 1\arcdeg{} image size does not capture the full field of view for the GLEAM sky-model, so some components do not appear in the image. However they will make a contribution to the visibilities. 
The choice of imager is arbitrary, as we compared the results from the compressed data to those from the non-compressed data, rather than the input catalogue. The cleaning parameters are also somewhat arbitrary given our comparison methodology.

The vital component for this work was the evaluation of the possible radio astronomy data degradation caused by lossy compression and its impact on image reconstruction. Common diagnostic tests in, say, the SKA Data Processing pipelines were deemed insufficient as they do not expose some of the artifacts that might be present in the resulting image.  These diagnostics measures include global root-mean-square (RMS) across the image, source positions, and source flux \citep[RASCIL]{rascil}. 
The RMS is a measure of the global image quality and the latter two diagnostics test the general quality of the corrections (in that poor calibration or poor apriori information, such as antenna positions or those that effect the reference frame, which will shift apparent positions and/or the coherence of the sum). The key limitation of these diagnostics is their inability to detect subtle effects that could be localised and perhaps associated with regions close to strong sources. 
For simulated data neither of the mentioned effects would apply, therefore we put together some additional alternative tests, which are more suitable for checking subtle image degradation. These can be combined with those tests of the calibration mentioned above to provide a complete test suite.

The metrics we studied are: image RMS, residual RMS, localised RMS, Kurtosis, two point correlation, the maximum and minimum values and these values over the RMS. 
For these investigations the image RMS is the RMS in the difference between images made with compressed and non-compressed data.
The residual RMS is the RMS in the residual images after deconvolution and model subtraction made with compressed and non-compressed data. 
The localised RMS is formed in sub-regions of the difference image (on a 32 by 32 grid in this case). 
The Kurtosis is formed from the second order moment on both the residual and the difference images.
The two point correlation is the maximum absolute pixel value in radial rings of the FFT of both the residual and the difference images.
The maximum and minimum are the largest and smallest values in the residual and the difference images.

The RMS and absolute-maximum of the difference between images formed from compressed and non-compressed data will test whether the compression of the data changes the science outputs in a detectable fashion, globally and locally respectively. 
This is predicated on the assumption that our imaging  has produced outputs that are not limited by some other systematic limitation, due to any subtle non-linear processes inherent in the imaging of radio-interferometry data. 
To minimise these contributions we have formed images with and without the CLEAN deconvolution, as CLEAN is widely considered to be the dominant non-linear process from the analysis \citep{clean-nonlinear}. 

The ratio of the two point maximum correlation of the images formed from compressed and non-compressed data is sensitive to subtle changes due to the compression that might introduce some bias in the average of the values. Such an effect would appear in the image, as every pixel in the image can be thought of as the phased, weighted sum of all the data in that direction. The most common introduced effects are ripples across the image due to the incorrect reconstruction of a source, with the unmodelled flux then `scattered' across the image. Additionally any introduced astrometric coordinate error would appear as localised ripples around the mismodelled sources.  

\section{Results and Discussion}
\subsection{MGARD Compression of different sky models}

\begin{figure}
    \centering
     \includegraphics[width=0.99\linewidth, clip, trim={0.cm 0.cm 1.8cm 0.cm}]{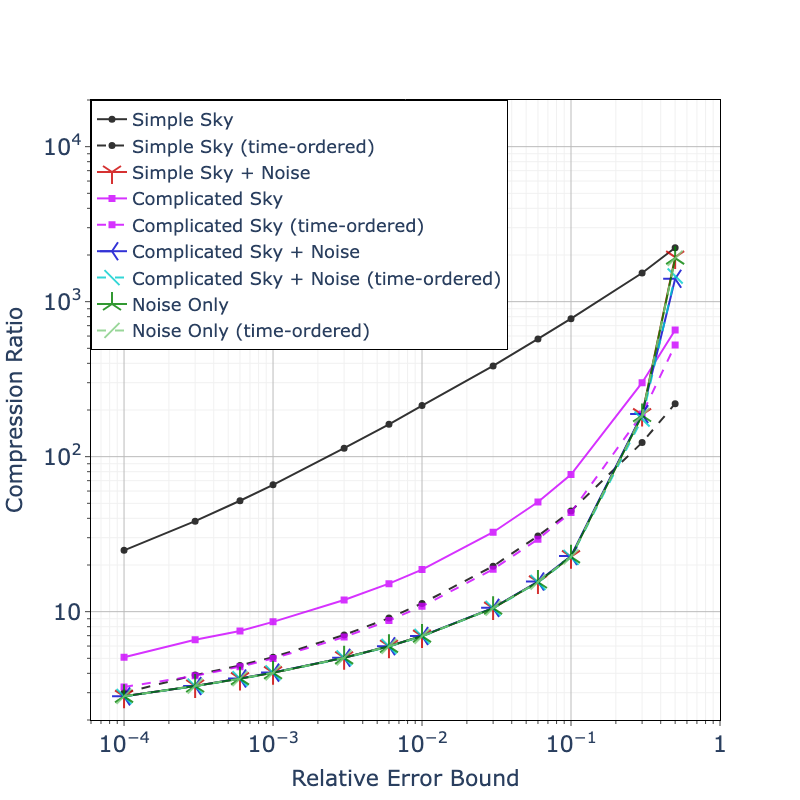}
    \caption{{The compression ratios achieved when compressing the simulated visibility data with various noise profiles. Shown are the compression results from MGARD for data with a complicated sky (purple squares) and a simple sky (black circles) model. 
    The x-axis describes the relative error bound provided to MGARD for the compression. 
    For the data that are dominated by thermal noise (line-style symbols, multiple colours) the compression ratios as a function of EB are practically identical and overlap on this plot.
    {The highest compression is achieved when the sky model is simple and noise free and ordered by baseline (black line). 
    The noise-free baseline-ordered complicated sky (purple line) is an order of magnitude lower, representing the impact of the rapidly changing sky signals. 
    Finally the time-ordered data is represented with dotted lines. The noise-free time-ordered data has significantly lower compression ratios than the baseline-ordered data, as the time-ordering hides the sky signal from the compression algorithm. In our simulations the complicated and simple sky in time-order provide very similar compression ratios for the same EB bound, particularly for the less aggressive compression, underlining the importance of presenting the sky signal coherently to the compression algorithm.  
    } 
    }}
    \label{fig:cr_eb_ri}
\end{figure}

We evaluated the compression ratio obtained on the data in the real-imaginary format for a range of simulated sky models. 
These were: the complex and realistic sky represented the GLEAM sky-model and a simplified compact model at the phase centre (where the model would change smoothly), both with and without added Gaussian random noise, as shown in Figure \ref{fig:cr_eb_ri}.
{In a subset of the tests this}
data was reorganised to be baseline ordered (i.e. time-fast), allowing the compression algorithm to `see' the baseline data where sky signals would be smoother. 
{Otherwise it was retained in time-ordered format; for noise-dominated data, for which every sample is independent, the ordering did not affect the compression ratio. 
}
The compression shown used MGARD with relative EBs between $10^{-4}$ -- $5\times10^{-1}$. 
Where the Gaussian noise dominates, for complicated, simple or noise-only data (blue, red and green), the compression ratio  varies from a factor of 3 at EB of 10$^{-4}$ to several thousands, with about a factor of 20 at an EB of 10$^{-1}$.
The complicated sky model without noise (purple) compressed several times better than the noise. 
The simple sky model without noise (black) compressed more than an order of magnitude better, with the compression ratio 
of about a factor of 800 at an EB of 10$^{-1}$. 
If the noise-free data was not reordered, so that it was in the default time-ordered format (dotted lines), the compression ratio fell by more than an order of magnitude, to be only few times better than noise dominated signals.
{
For the noise-dominated signals the ordering or the background signal did not matter and all compression ratios were very similar for a given EB.
This is inline with expectations, as smooth data will compress best and baseline ordered data, without noise and with a simple sky model, will present the smoothest data to the compression algorithms.
}
Thus we conclude in a noise-free situation reordering before compression could have an impact, 
however, as the there are only a few places in the sky where a single simple source dominates over the noise, this will not be a widely applicable domain.
As this would be an expensive and I/O intensive operation we would not recommend it in general,
{however it would be an option for strong and simple sources, such as the calibrator scans}. 

\subsection{MGARD Compression of amplitude and phase}

\begin{figure}
    \centering
    \includegraphics[width=0.99\linewidth, clip, trim={2.cm 0.cm 1.cm 0.cm}]{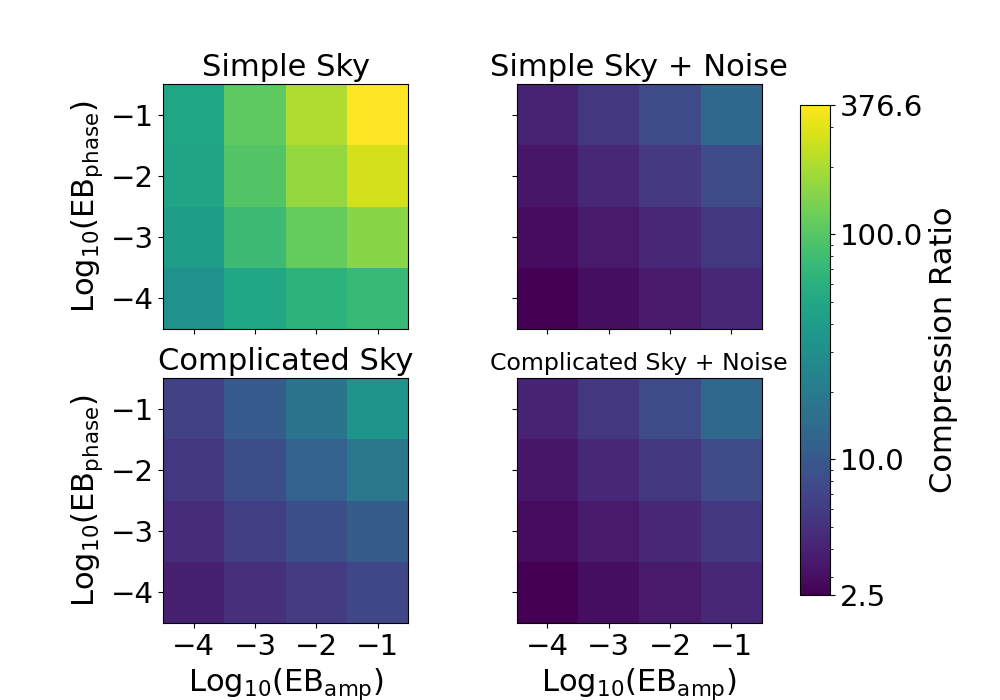}
    \caption{The compression ratios achieved when compressing the amplitude and phase of the visibility data. Shown are the results for simple sky simulation (top), a complicated sky simulation (bottom), both with (right) and without (left) added noise. 
    The x and y-axes are the log of the relative error bounds provided to MGARD individually for the amplitude and phase. 
    }
    \label{fig:cr_eb_ap}
\end{figure}

To further explore expressions of the data where the signal might be smoother and therefore compress better, we investigated the compression ratio for cases were the amplitude and the phase were compressed separately, with independent EBs. 
These stepped from $10^{-4}$ to $10^{-1}$ in steps of ten, as shown in Figure \ref{fig:cr_eb_ap}.
The compression ratio was symmetric around the axis of EB$_{\rm phase}$ equal EB$_{\rm amp}$.
{We found that compressing in amplitude and phase only provided an advantage for simple model in noise-free domain, at low compression. For the noise-dominated sky-models, and for higher compression ratios, 
compressing the real and imaginary data gave about a 20\% better performance in all regimes of relevance, i.e. below an EB of 0.1. }
Thus we find no advantage of converting from the natural real and imaging axis, 
nor in reordering the data to expose any potential smoothness.
For this reason for MGARD we would not recommend compressing the complex data in amplitude and phase components.

\subsection{Comparison of MGARD Compression with DYSCO Compression}





\begin{figure}
    \centering
    \includegraphics[width=0.99\linewidth, clip, trim={0.cm 0.cm 1.8cm 0.cm}]{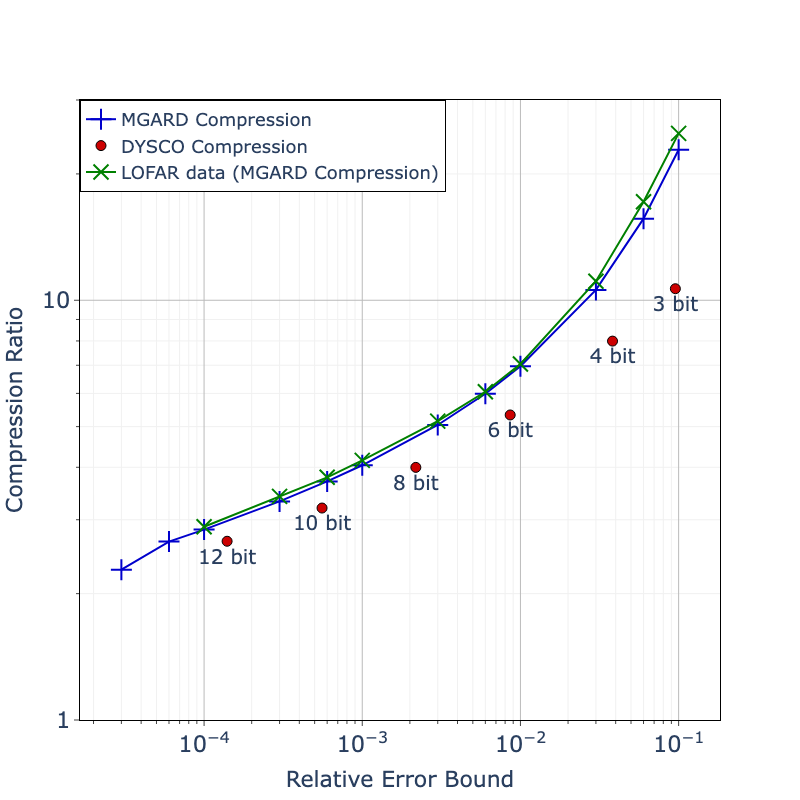}
    \caption{
    A comparison {in log-log} scale between the compression performance of MGARD and DYSCO for the time-ordered ``Complicated Sky + Noise'' case (compare with {cyan} line in Figure \ref{fig:cr_eb_ri}). The compression parameters for DYSCO are defined in terms of bit storage as indicated and so the shown relative error bounds have been calculated from the 99\% percentile of the residuals in order to conform with those in MGARD. In addition we plot the actual complicated sky and noise from the MGARD compression of LOFAR observations, where we have used absolute error bounds and estimate the equivalent relative error bound for compression assuming a data range of $\pm$5$\sigma$. 
    }
    \label{fig:dyscoVmgard}
\end{figure}

{This comparison was performed entirely on} time-ordered data, as DYSCO did not support the baseline-ordered format. For the simulated data we used the time-ordered complicated sky-model with noise, which can be compared to the {cyan} line in Figure \ref{fig:cr_eb_ri}. 
The DYSCO compression ratios are by bit reduction, with fixed factors of 2, 3, 4, 6, 8, 10, 12 or 16, although we note that 8-bit and above are the recommended ranges \citep{dysco}. 
We used the 99\% percentile as the approximate equivalent error for DYSCO. 
This was used to compare with the  requested error bound pre-compression in MGARD, as shown in Fig. \ref{fig:dyscoVmgard}.
For an MGARD relative EB above $10^{-4}$ the data compression ratio is larger than a factor of 3, and a factor of 4 at $10^{-3}$ and a factor of 7 at $10^{-2}$.
Above an EB of 0.1 the compression ratios are greater than twenty, as the data becomes highly quantised with all the data compatible with a few values across the whole dataset. 
{For the DYSCO compression the performance is slightly worse than for MGARD, with the compression ratios falling below the MGARD for the equivalent error by 50\% at 3-bit compression and improving to $\sim$10\% at 12-bit compression. 
{However, we caution that this is only an estimate, particularly under aggressive compression, as it is not an direct comparison of true like-for-like error estimates.}
{Furthermore, DYSCO uses a non-linear quantisation format, which allows for recovery of this shortfall in the imaging tests.}
}
For the LOFAR compression we could not use use the relative EB, as we can not use the actual data range, as it is highly non-Gaussian with large tails of a few outlying values due to RFI. {
This allows us to test the performance of MGARD in the face of real-life non-ideal behaviour.
We found that setting the absolute bounds to match the expectations based on the thermal noise, or alternatively the standard deviation, would deliver performance close to that of perfect Gaussian distributed data, as shown by the close correspondence of the LOFAR and simulated data results.}
We used absolute EBs between 100 and 2E-3\,Jy, and for plotting on Fig. \ref{fig:dyscoVmgard} and assume that the data range should be $\pm$5$\sigma$ (i.e. $\pm$20\,Jy) {for the conversion to a relative EB}. 
In this case the compression ratio tracks the results from the simulations, giving us confidence in our conversion from absolute to relative EB.

{The other strong advantage that MGARD offers in comparison to DYSCO is the precisely tunable selection of the error bounds, and thus the compression ratio. 
MGARD is much more flexible than the fixed levels of compression offered by DYSCO, and furthermore the adjustable parameter can be directly matched to the system noise level.}

{ 
A future investigation that merits exploration is to utilise the MGARD capability to control the compression over regions of interest, and use this to vary the error bounds as a function of baseline length. This would use the multi-grid approach to form baseline dependent averaging, which we would expect to give a more precise reconstruction than simply averaging data into the equivalent of uv-cells. For the moment we are attempting to surpass this by the compression of the uv-grids themselves \citep{williamson-24,williamson-25} where the averaging is formed after the correct application of the weighting kernels, but a comparison would be interesting.}

\subsection{Image Quality Analysis}

\begin{figure}
    \centering
    \includegraphics[width=0.99\textwidth, clip, trim={0.cm 0.cm 0.5cm 0.cm}]{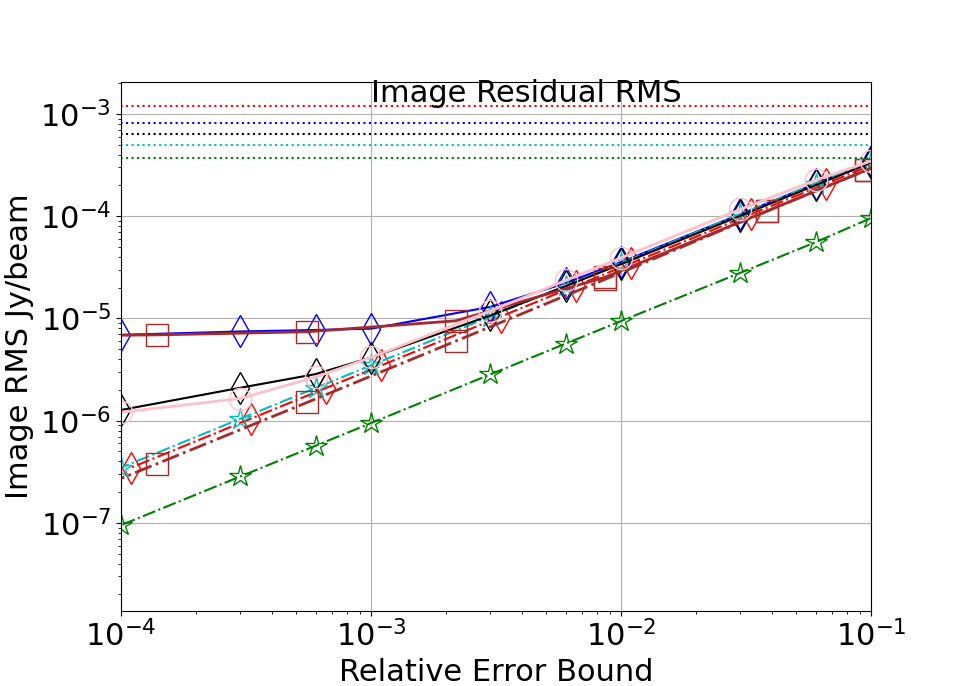}
    \caption{RMS of the difference between the image from compressed data and the non-compressed data;
    noise-free data is marked with a star and noise-added data is marked with a diamond. 
    Some markers are slightly shifted for clarity.
    The results from cleaned images are shown with a solid line; a dot-dash line is for non-cleaned images.
    The MGARD compression impact on complicated sky-images, with and without noise, and without and with deconvolution are shown in red, blue and cyan respectively. 
    The simple and high dynamic-range image with noise and cleaning is shown in black, and without noise and cleaning is in green.
    DYSCO compressed data with and without cleaning is shown in purple with squares. 
    An example for real data, cleaned and with intrinsic noise, from the LOFAR pathfinder is shown in yellow.
    The lighter dotted lines without symbols represents the RMS of the residual images made from non-compressed data. 
    One can directly see that the global added noise to the images from using compressed data is significantly less than the image noise. 
    Furthermore, one can see that non-linear process of cleaning on the GLEAM-model (which can be imaged to a dynamic range of about 1,000 with 64 antennas) produces a limit on the reconstruction precision. That is reducing the degree of compression, and thus precision, does not improve the reconstruction accuracy. 
    However for the higher dynamic range ($>$10,000) images the reconstruction accuracy continues to improve with the improved precision.
    The simulated DYSCO-compressed and real MGARD-compressed data have similar results to the simulated MGARD-compressed data.
    }
    \label{fig:rms}
\end{figure}

The RMS of the image difference in a pixel to pixel comparison between images formed from compressed and non-compressed data is shown in Fig. \ref{fig:rms}. 
MGARD-compressed simulated data dominated by signal (noise-free) is marked with a star and the data dominated by noise (noise-added) is marked with a diamond. 
{The real LOFAR data is marked in solid pink lines with circles. 
The DYSCO-compressed data is marked in purple with squares.}
Solid lines indicate images that have been deconvolved with CLEAN and dot-dashed lines indicate where no deconvolution was performed.
The RMS of the images made with the non-compressed data is shown as the dotted horizontal lines at the top of the plot in the same colours. 
The RMS of the difference between non-compressed and compressed data images falls significantly below the image residual RMS, indicating that the added noise from the compression is much less than that in the images of the non-compressed data itself. 
We note the dominant impact is actually in the deconvolved images, where the decovolution reaches an accuracy limit and improved precision in the compression does not improve the reconstruction. 
For the complicated sky, where the limited number of antennas limit the possible accuracy to about a dynamic range (DR) of 2,000 we see no improvement below an EB of $10^{-3}$. For the simpler sky model, where we can achieve a DR of 14,000, the reconstruction continues to improve beyond this limit, albeit at a lower rate.
{The DYSCO points on this plot are slight to the right of the of the equivalent MGARD points, implying that they achieve the same recovered image quality for a larger relative error; this is presumably, as all other aspects are equal, due to the non-linear quantisation. However these gains (approximately improving the performance by a factor of 1.04) are nullified by the better compression for MGARD.}
{The real LOFAR data has results very similar to those from the simulations, particularly if we bear in mind the uncertainties in the conversion of absolute to relative EB for the LOFAR data.}

\begin{figure}
    \centering
    \includegraphics[width=0.99\textwidth, clip, trim={0.cm 0.cm 0.5cm 0.cm}]{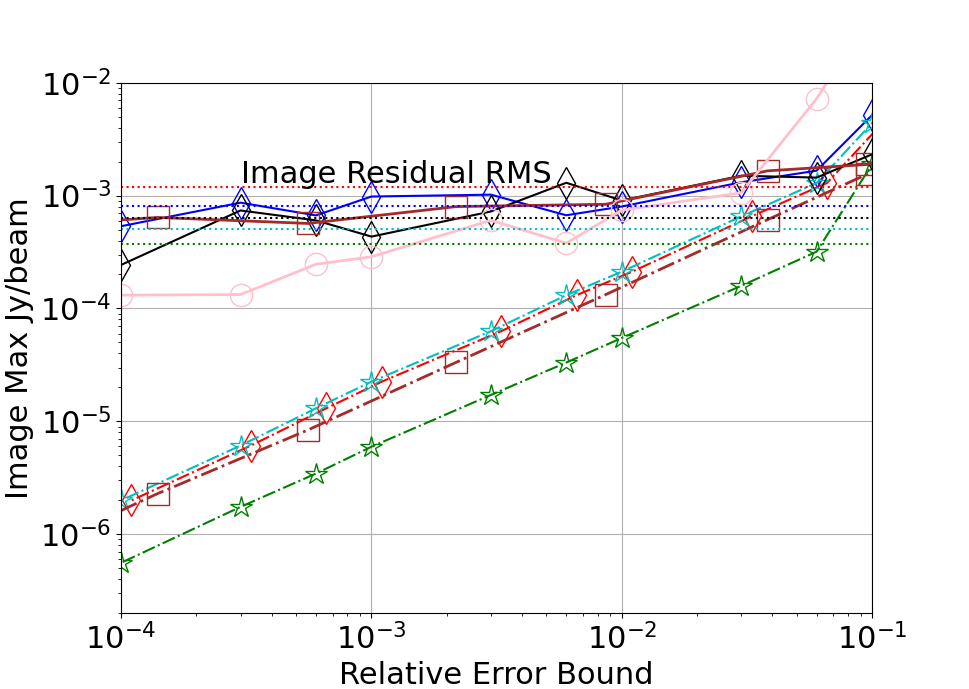}
    \caption{Maximum value of the difference between the image from compressed data and the non-compressed data, with the same data labelling as Fig. \ref{fig:rms}, {including the dotted lines without symbols indicating the RMS of the residual images, for comparison.}
    The maximum error between the images of compressed and non-compressed data picks up (predominately) the change in side-lobes around the strongest sources. Thus, if the image in CLEANed, the values tend to {those of} the image RMS. 
    For MGARD, with high compression ($>10\%$), the images of compressed data started to significantly diverge from the non-compressed images, whilst DYSCO continues to perform well. This may, however, be more related to the poorer compression ratio of DYSCO.
    } 
    \label{fig:max}
\end{figure}

The RMS only measures the global difference, to investigate the impact of differences localised in the image we repeated the analysis above, but for the absolute maximum. 
{These results are shown in Fig. \ref{fig:max} with the same colours as in Fig. \ref{fig:rms}.
Here we see how the deconvolution mixes the thermal noise with the compression errors so that the maximum errors track the image RMS. The cause of this effect is clearly notable in the sidelobes around the strongest sources in the image, which blend the two noise sources to produce thermal level variations that are a function of the strength of deconvolved sources.
This effect is not seen in the images without CLEAN applied, and presumably any other deconvolution method that did not suffer from the CLEAN instabilities.
}
%

One should bare in mind that SKA images could be stacked to achieve greater sensitivity. That is multiple independent images would be combined to make a deeper image.
Provided the compression does not introduce systematic effects, which should be the case where it is dominated by thermal noise, we would expect the compression noise to also average down. However, in the thermal noise-free case one could imagine a case where the compression noise could become the systematic limit. 
We would recommend compression error bounds of below $\lesssim10^{-3}$ to ensure an image error bound of a few $\mu$Jy, for the given simulation parameters. This is a factor of a hundred less in RMS than the intrinsic error bound and could only dominate if more than 10,000 individual images were combined.


\begin{figure}
    \centering
    \includegraphics[width=0.99\textwidth, clip, trim={0.cm 0.cm 1.cm 0.cm}]{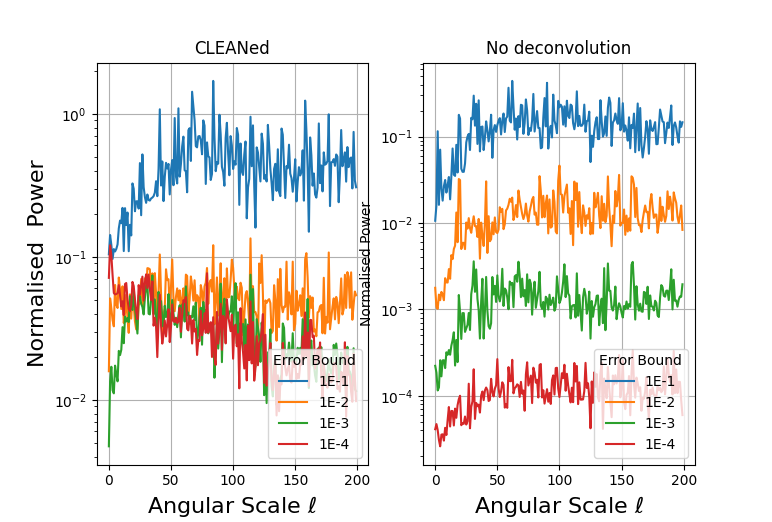}
    \caption{Two point correlation of the images showing the maximum power at all angular scales ($\ell$ from zero to 200, where zero would be a constant value across the $\sim$1\arcdeg{} image and 200 would be the power on 16$^{\prime \prime}$ scales) from the cleaned images made with compressed data, normalised by the power in the image made with non-compressed data. 
    On the left is shown the results for the complicated GLEAM-model, where the reconstruction accuracy reaches a limit at an error bound of $\sim$1E-3. 
    On the right is shown the result for the same data, but imaged without deconvolution, which does not introduce a limit in the reconstruction accuracy, which continues to improve with EB precision.
    No peak from introduced image artefacts are detectable, and all contributions are well below the noise levels.
    }
    \label{fig:ell}
\end{figure}

The ratio of the two point maximum correlation of the images formed from compressed and non-compressed data with a complicated sky model is shown in Fig.  \ref{fig:ell}.
To the left the deconvolved data is shown, where the CLEAN step introduces a limit in the achieved precision of the reconstruction. 
The right shows the results from the same data without deconvolution.
The profiles are close to white noise (i.e. flat), that is there are no detectable large scale ripples or similar features.
This is because MGARD transforms data into small-valued coefficients through multi-linear interpolation then performs quantisation on interpolation residuals. 
Given the original data is overwhelmed by thermal noise, the interpolation residuals will exhibit a random distribution akin to the white noise. The quantisation will then introduce an almost \textit{uniform} loss across the entire data space. 
The gradient at large angular scales we interpret as the data compression preserving the large scale structure slightly better than the small scale structure. Nevertheless, the limits are well below the native image reconstruction errors, even in the noise-free case.

\subsection{Effect of compression on spectral line cubes}


\begin{figure}
    \centering
    \includegraphics[width=0.99\linewidth, clip, trim={0.cm 0.cm 2.cm 0.cm}]{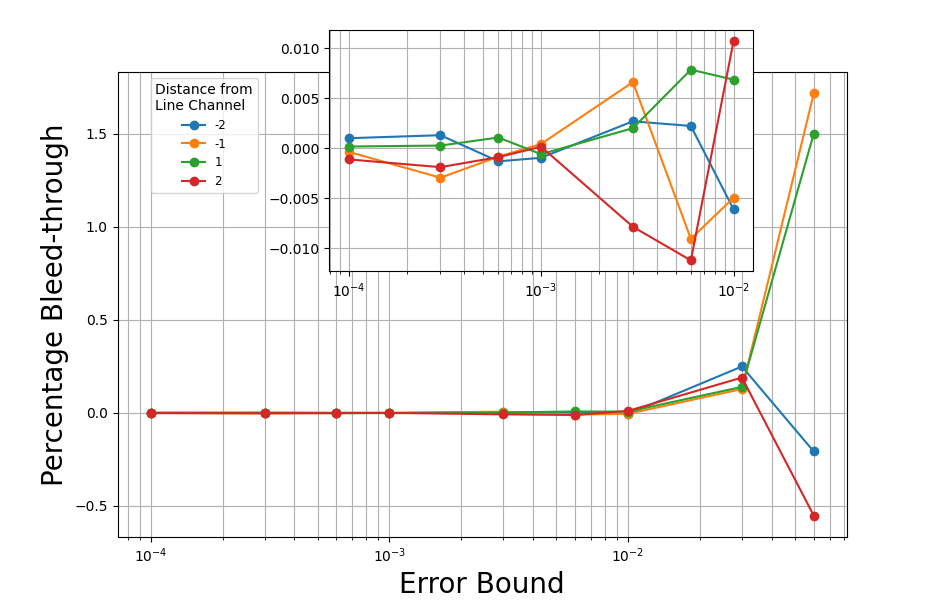}
    \caption{The degree of bleed-through from the strong spectral feature into surrounding channels, as a function of the compression error bound. As would be expected the cruder compression causes greater bleed-through, although still at less than 1\%. From an error bound of 1E-2 or below the bleed through is less than a few times a factor of 1E-5.}
    \label{fig:spectral-bleed}
\end{figure}

To test the impact of the compression on the well-known `bleeding through' of the strong spectral feature into the surrounding channels we added this simple sky point source to one channel of the visibilities then compressed the whole dataset and imaged the data, channel by channel. 
Figure \ref{fig:spectral-bleed} plots the peak flux in the surrounding channels, normalised by the peak flux of the spectral feature (28.5Jy/beam). Even for a relative error bound below 10$^{-1}$ the fractional error is the order of a percent, for error bounds of $10^{-2}$ and below the fractional error is about $10^{-4}$.
We note that our implementation in ASKAPSoft, as it performs the MGARD compression in parallel on a (independent) channel by channel basis, will not suffer from this effect. 
{However, these results will impact other future potential implementations of MGARD and thus are included for completeness.}

\section{Conclusions and Outlook}

\subsection{Conclusions}

We have demonstrated the functionality of the MGARD compression application on the complex visibilities one might expect from the next generation of instruments. 
This was on both simulated SKA data and real pathfinder data from LOFAR. 
MGARD matches or slightly exceeds the performance of the best current option for data compression, whilst providing a much more natural and flexible data compression metric.
The concept of an error bound allows us to guarantee that the data is not degraded more than the specified noise levels.
The compression of the data, without loss of information, directly addresses one of the major cost drivers of the SKA.

Selecting relative error bounds less than 10\%, with a compression ratio of 20, introduced no significant errors in the continuum imaging, whereas 1\% error bound, with a compression ratio of about 8, introduced a loss of precision about an order of magnitude less than the noise. Below 0.1\% (compression ratio of 4) imperfections in the deconvolution dominated. 
For the spectral line imaging, EBs below 10\% limited the bleed through to 2\% or less of the peak flux, whereas a 1\% EB limited the bleed through to less than 0.03\% in the worst case. 
In summary, we find that EBs between $10^{-3}$ and $10^{-2}$  have a good compromise between compression and impact on the science data products. 
{High compression ratios achieved with small information loss suggest that the raw data was significantly oversampled. This is consistent with the results we have obtained, as 
in the signal-dominated case a smoothly changing signal is highly compressible 
using the recommended error bounds (i.e. 1\% of the range).
In the noise-dominated case we obtain more moderate compression but still up to an order of magnitude of improvement over the non-compressed data. This is because there was no useful information significantly below the resolution imposed by the thermal noise.
}

We did not find any benefit from reordering the data except for the simplest noise-free cases, hence we believe the best recommendation for the use of MGARD on the SKA would be to compress the data from the correlator on the fly to an error bound of 0.1\%. 
This could quarter the short term buffer storage costs and massively ease the pressure on the SKA budget. 

\subsection{Outlook}
MGARD offers a host of additional features which may have great applications in Radio Astronomy. 
{It continues to be improved in performance and, as it is now being optimised specifically for Radio Astronomy data, we expect further increases in compression ratios achieved.}
{The innovative non-linear quantisation as used by DYSCO can be implemented in MGARD to reduced the image reconstruction errors for a given compression level.}
As MGARD is embedded in ADIOS2 one also has direct access to highly parallelise reading and writing, reducing I/O requirements.
{
Recent studies \citep{Gong:2023} demonstrate that MGARD achieves a throughput of 15GB/s and 30GB/s when compressing data using an NVIDIA A100 and AMD MI250X GPU. An evaluation by \cite{Chen:2025}, using 1,024 nodes on Frontier supercomputer show that MGARD can accelerate the write and read operation of ADIOS2 by $6.8-15.3\times$ and $5.2-9.3\times$, with $14-2379\times$ compression ratios.}
{Our next paper \citep{williamson-25} discuss the performance on MGARD in ASKAPSoft on DINGO in detail.}

{We have not used the rich feature-set of MGARD in this work.
For example, one could be particularly interested in the low or high resolution data. MGARD allows for preferential preservation of precision based on regions, such as baseline length. 
RoIs could be used to implement an optimal Baseline Dependent Averaging (BDA) approach as, in the gridding step, the short baselines are more heavily averaged in the imaging. 
Thus this feature could used to implement increased averaging in the shorter baseline visibilities with controlled precision before the gridding, to increase the compression achieved. 
As MGARD uses a multi-grid method this should out-perform the traditional BDA compression which simply step-wise averages over variable time intervals that depend on the baseline length.
}

MGARD can also be applied to other data products in the SKA data lifecycle. We have already investigated the applicability of MGARD to gridded visibilities from ASKAP \citep{williamson-24,williamson-25}. Similar to \citet{Kitaeff:2015} and \citep{Peters:2014} we are also planning to investigate MGARD's flexible capabilities, like hierarchical compression and special treatment of regions of interest (RoIs), to compress radio astronomy image cubes. The results of these investigations have the potential of very significant cost benefits for the SKA project as a whole in a range of areas from intermediate storage and I/O, over LAN and WAN network costs to archival storage. In some cases applying compression might even enable certain science projects, which would otherwise be unfeasible due to data volume or I/O constraints.


\begin{acknowledgement}

AW acknowledge the Pawsey Centre for Extreme Scale Readiness (PaCER) for funding and support. 
This research has used the facilities of the Pawsey Supercomputing Research Centre, The Oak Ridge National laboratory, the LOFAR archive and the OSKAR software. We would like to acknowledge the Whadjuk people of the Noongar nation as the traditional custodians of this country, where the Pawsey Supercomputing Research Centre and ICRAR is located. We pay our respects to Noongar elders past, present, and emerging. 
PJE and AW would like to acknowledge the many invaluable conversations with the ASKAPsoft development team.
\end{acknowledgement}

\paragraph{Funding Statement}

This manuscript has been authored in part by UT-Battelle, LLC, under contract DE-AC05-00OR22725 with the US Department of Energy (DOE). This research was supported by the SIRIUS-2 ASCR research project, the Scientific Discovery through Advanced Computing (SciDAC) program, specifically the RAPIDS-2 SciDAC institute, and the GE-ORNL CRADA data reduction project. 


\paragraph{Competing Interests}

None

\paragraph{Data Availability Statement}

Our Data Quality Analysis algorithms are publicly available on github.com/ICRAR/Image-DQA/. 
The LOFAR data was taken from the long-term archive lta.lofar.eu/. 


\printendnotes
\printbibliography



\end{document}